\documentclass[letterpaper,10pt,twocolumn,final,journal,oneside]{IEEEtran}

\usepackage{multirow}
\usepackage{graphicx}
\usepackage{acronym}
\usepackage{amsmath}
\usepackage{eurosym}
\usepackage{siunitx}
\usepackage[caption=false,font=footnotesize]{subfig}
\usepackage{stfloats}
\usepackage{tikz}
\usepackage{hyperref}
\usepackage{ccicons}
\usepackage{balance}
\usepackage{xcolor}

\acrodef{nra}[NRA]{national regulatory authority}
\acrodefplural{nra}{national regulation authorities}
\acrodef{lv}[LV]{low voltage}
\acrodef{sm}[SM]{smart meter}
\acrodef{dso}[DSO]{distribution system operator}
\acrodef{tso}[TSO]{transmission system operator}
\acrodef{plc}[PLC]{power line communication}
\acrodef{rf}[RF]{radio frequency}
\acrodef{esco}[ESCO]{energy service company}
\acrodefplural{esco}{energy service companies}
\acrodef{pod}[POD]{point of delivery}
\acrodef{bsp}[BSP]{balancing service provider}
\acrodef{mevu}[MEVU]{mixed enabled virtual unit}
\acrodef{ami}[AMI]{advanced metering infrastructure}
\acrodef{ev}[EV]{electric vehicle}
\acrodef{han}[HAN]{home area network}
\acrodef{hems}[HEMS]{home energy management system}

\def \chaintwo {\textsc{chain~2}}
\def \chainone {\textsc{chain~1}}

\DeclareSIUnit{\kWh}{kWh}
\newcommand\copyrighttext{%
  \centering\ccbyncnd\\ \copyright~2021. This manuscript version is made available under the\\
  \href{http://creativecommons.org/licenses/by-nc-nd/4.0/}{Creative Commons Attribution-NonCommercial-NoDerivatives 4.0 International License.}\\
  Applied Energy, vol.~304, 15 Dec. 2021, article no.~117806 DOI: \href{https://doi.org/10.1016/j.apenergy.2021.117806}{10.1016/j.apenergy.2021.117806}.}
\newcommand\copyrightnotice{%
\begin{tikzpicture}[remember picture,overlay]
\node[anchor=north,yshift=0pt] at (current page.north) {\setlength{\fboxrule}{0pt}\fbox{\parbox{\dimexpr\textwidth-\fboxsep-\fboxrule\relax}{\copyrighttext}}};
\end{tikzpicture}%
}

\begin{document}

\title{Post-Metering Value-Added Services for Low Voltage Electricity Users: Lessons Learned From the Italian Experience of \textsc{chain~2}}

\author{Daniele~Serra, Daniele~Mardero, Luca~Di~Stefano, and Samuele~Grillo%
\thanks{D. Serra, D. Mardero, and L. Di Stefano are with E-Distribuzione S.p.A, via Ombrone, 2 00198 Roma, Italy (e-mail: \{daniele.serra, daniele.mardero, luca.distefano\}@e-distribuzione.com).}%
\thanks{S. Grillo, is with the Dipartimento di Elettronica, Informazione e Bioingegneria, %
Politecnico di Milano, piazza Leonardo da Vinci, 32, I-20133 Milano, Italy %
(e-mail: samuele.grillo@polimi.it).}%
}

\IEEEaftertitletext{\copyrightnotice}
\maketitle
%\IEEEpeerreviewmaketitle

\begin{abstract}%\linenumbers
Electrical energy consumption data accessibility for low voltage end users is one of the pillars of smart grids. In some countries, despite the presence of smart meters, a fragmentary data availability and/or the lack of standardization hinders the creation of post-metering value-added services and confines such innovative solutions to the prototypal and experimental level. We take inspiration from the technology adopted in Italy, where the national regulatory authority actively supported the definition of a solution agreed upon by all the involved stakeholders. In this context, smart meters are enabled to convey data to low voltage end users through a power line communication channel (\chaintwo) in near real-time. The aim of this paper is twofold. On the one hand, it describes the proof of concept that the channel underwent and its subsequent validation (with performances nearing 99\% success rate). On the other hand, it defines a classification framework ($\rm I^2MA$) for post-metering value-added services, in order to categorize each use case based on both level of service and expected benefits, and understand its maturity level. As an example, we apply the methodology to the 16 use cases defined in Italy. The lessons learned from the regulatory, technological, and functional approach of the Italian experience bring us to the provision of recommendations for researchers and industry experts. In particular, we argue that a well-functioning post-metering value-added services' market can flourish when: i) distribution system operators certify the measurements coming from smart meters; ii) national regulatory authorities support the technological innovation needed for setting up this market; and iii) service providers create customer-oriented solutions based on smart meters' data.
\end{abstract}


%\linenumbers
%===============================================================================
\acresetall
\section{Introduction}

\IEEEPARstart{O}{ver} the last decades, electricity has proved to be not only an efficient, sustainable and widespread energy vector, but also natively inclined to adapt and incorporate new technologies. One of the major challenges the electrical energy sector is facing nowadays is how to combine digitalization with the ongoing evolution of the grids towards a smart paradigm~\cite{Farhangi2010}, which involves the incorporation of new devices~\cite{Lopes2020} and a general necessity to rethink the structure of transmission and distribution grids~\cite{Razon2020,Vadari2020}. The new opportunities provided by the automation of power systems and the improvement in the quality of service have contributed to create a perception of electrical energy as a stable, certain, and unlimited commodity, especially in developed countries. Only recently has the final user realized that more rational usage of electricity is necessary for the purposes of a sustainable and inclusive development~\cite{Maheswaran:2012,Bazydlo:2018}, as defined by the United Nations' Agenda 2030~\cite{UN2030}.

Since one of the best ways to attain this is to make the ``last mile'' more efficient~\cite{Anda2014,Carroll2013}, having a good measuring equipment is vital in order to estimate and understand the beneficial effects of this approach~\cite[p.~354]{Dobre:2016}, \cite{Corbett:2013}. On the other hand, the grid is in need of reliable and frequent consumption and production data from customers in order to be able to cope with the challenges introduced by the integration of distributed energy resources into power systems~\cite{IRENA:2021,Sinsel:2020}, as well as the creation of robust electric vehicle charging infrastructures~\cite{IEA:2020,Kapustin:2020}. With the goal of offering more flexibility to the grid and adapting to renewable sources' production patterns, new opportunities must be studied, such as demand-side management~\cite{Palensky2011,Sarker:2021} and local flexibility markets~\cite{Jin2020}. The presence of \ac{ami}, which comprises \acp{sm}, communication networks, and data management systems, proves to be a powerful tool and offer a significant contribution to both these objectives~\cite{Mou2018,RashedMohassel:2014}. In this regard, the role of the \ac{dso} is rapidly changing, as pointed out by IRENA~\cite{IRENA:2019}.

So far, many have studied technological, functional and behavioral factors affecting the use of SMs' data both on the grid side and on the user side (see, for instance, the case studies described in~\cite{Fadel:2015}, \cite{Volker:2021} and \cite{Zhou:2016} respectively, and refer also to Section~3). To the best of the authors' knowledge, nobody has coupled this with the market structure and the regulatory approach, which can drive the creation of new technology-enabled services and transform already existing experiments into consolidated cases. The liberalization of electricity markets is a process that has affected most Western countries over the past decades~\cite{IEA:2005,Pepermans:2019}. In particular, starting from the end of last century, the European Commission and Parliament adopted several directives that drove the creation of competitive energy markets across the European Union~\cite{EPEC:1996,EPEC:2003,EPEC:2009}. Despite the persistent differences among countries and the fragmentation of policies, the main principle that leads liberalization, i.e., unbundling of production, distribution and trading activities, is the path that most policy makers have chosen to follow.
However, within this context, the regulation-driven adoption of a \ac{sm} technology has sometimes overlooked crucial aspects with respect to the possibility of enabling customer-oriented services (see Section~3 for further details). This article aims to understand the optimal way of enabling such near real-time, post-metering value-added services. It does so by analyzing the Italian experience, which has attained a high level of maturity and can provide a toolkit for the deployment and scale-up of similar solutions in other liberalized markets. Albeit local, the selection process for \chaintwo\@ technology and the associated policy implications do represent valuable guidelines for relatable cases.

The main contributions of this article are: i) a comprehensive definition of post-metering services' environment (Section~II); ii) an outline of their international framework and a literature review of the most relevant case studies (Section~III); iii) the description of a regulatory, technological, and functional methodology (called $\rm I^2MA$) for the analysis of post-metering services, their application to the Italian case (\chaintwo), and their significance for the development of an enabling infrastructure (Section~IV); and iv) the description of the lessons learned and the provision of recommendations for researchers, industry experts, and policy makers (Section~V). Finally, conclusions are drawn in Section~VI.

\section{Definition of Post-Metering Services' Environment}\label{sec:def}

Making measurements available to end users is one of the challenging aspects that smart grids represent for \acp{dso}. The problem to be tackled is not only to give customers access to data, but also, and mainly, to let them take full advantage of such data for real-time services. The paradigm change happens when measurements cease to be managed exclusively by the \ac{dso} internally and are delivered unaltered to end users externally. The availability of raw measurements frees customers from viewing their own data through the \ac{dso}'s lenses and paves the way to new value-added services' markets~\cite{VanAubel:2019,EG1:2015}.

% It is therefore useful to introduce the concept of post-metering services as the definition of the above-mentioned value-added services. %{\color{red}Post-metering services cannot exist without an \ac{ami}, which, in turn, comes into being through the deployment and exploitation of \acp{sm}.}

During the development of the first generations of \acp{sm}---which began in the early 2000s in some pioneer countries---the main concern was to guarantee easy access for \acp{dso} to remotely read consumption measurements and manage them, thus reducing manual intervention by personnel and, most importantly, enabling timely and effective responses to customers' or operators' requests. Thus, an electricity meter was defined ``smart'' or ``advanced'' when it could send its readings to the headend, mainly for billing purposes, and receive and process basic commands, such as enabling/disabling of the service and integrity checks~\cite{Owen:2006}. This paradigm can be regarded as the opening of a unidirectional communication channel from the meter towards the ``inner environment'', i.e., the \ac{dso} and the other entities that produce, dispatch, and sell electricity.

\begin{figure}[t]
	\centering
    \includegraphics[width=\columnwidth]{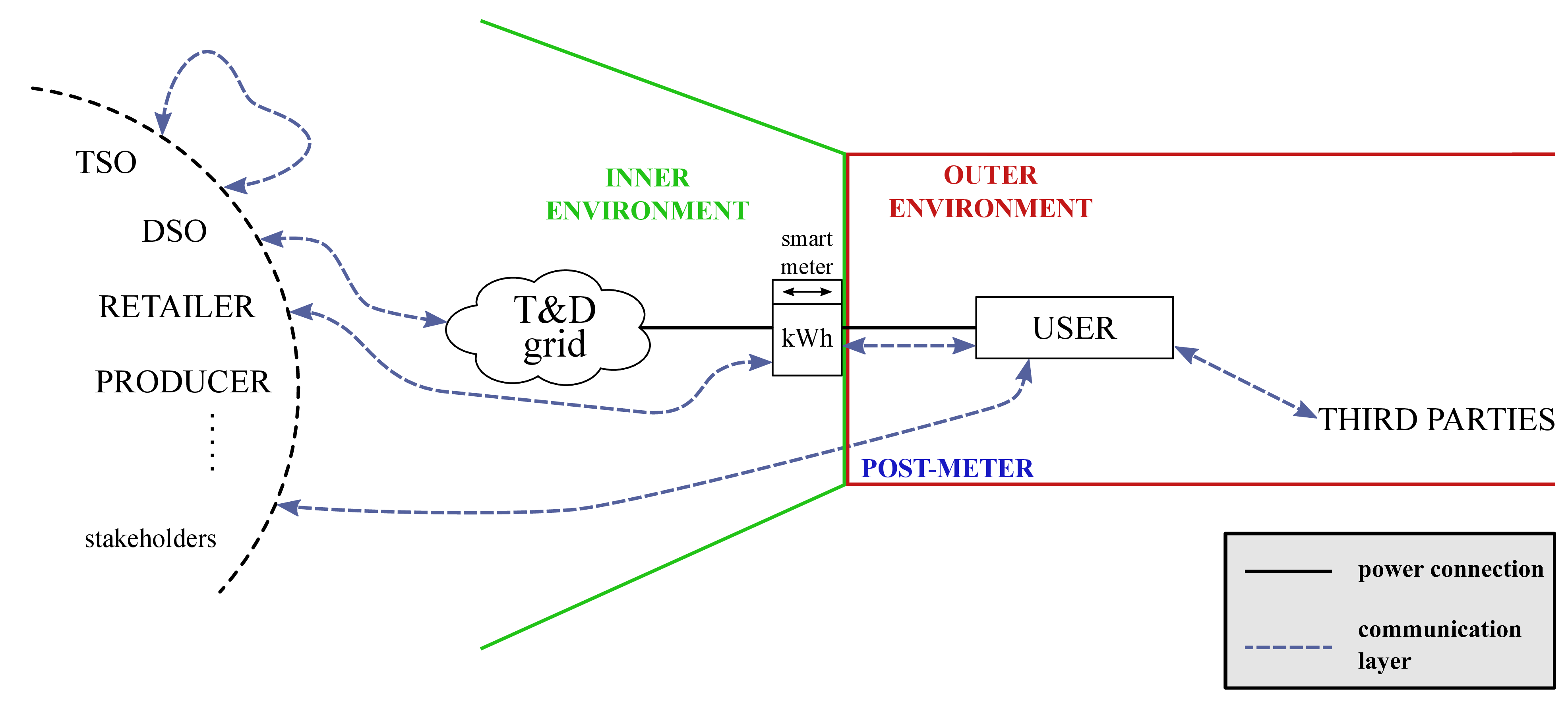}
    	\caption{Schematic description of the distribution system. The diagram (which is clearly not exhaustive) aims at defining the scope of the different stakeholders as perceived by the end user. The \ac{sm} is the boundary between the ``inner'' and the ``outer'' environments.}
	\label{fig:fig_1}
\end{figure}

Only more recently has the focus on \acp{sm} been put on the capability to communicate with the ``outer environment'', i.e., end users and the involved third parties (see Fig.~\ref{fig:fig_1} for a graphical representation of the classification of the inner and outer environments).
The first approach was to make the readings accessible to the end user locally---mainly by visual inspection---and remotely. The main purpose for doing this was to obtain reports \textit{a posteriori} as a tool to increase energy awareness and foster savings, and data were updated with considerable delay. After the first experimental trials and rollouts, \acp{dso} and researchers have come to realize this option is not sufficient to modify energy consumption patterns~\cite{ACEEE:2010}, especially when feedback is provided in an aggregated form~\cite{CarrieArmel:2013}, or with an unsatisfactory frequency~\cite{Fischer:2008,Chen:2014}. The key, instead, is to make the \ac{sm} a gateway through which end users are able to become active players in the electricity market, by connecting the devices installed on their own side and orchestrating their usage according to the \ac{sm}'s readings. This paper focuses on customer-oriented value-added services that come into being by opening a bidirectional channel between the \ac{sm} and the outer environment. These services are technology-enabled solutions that empower users to support the needs of the grid and achieve energy savings, thus providing the framework for new sustainable energy systems. It is therefore useful to introduce the concept of post-metering value-added services as the definition of such solutions. It is worth highlighting that the expression ``post-metering'' is a reference to their technical and architectural characteristics, while ``value-added'' underlines their commercial value. In this work we will use both these wordings, although not necessarily in conjunction.

There are some conditions which make a meter suited for this task. The three most important ones, which can be regarded as the ``pillars'' of post-metering services, are the following:

\subsubsection*{Data availability}
As already mentioned, post-metering services rely on the ability of the \ac{sm} to open data to the outer environment. The way this goal is achieved makes the difference between a merely notional data availability and an effective and actual availability. The two main characteristics of data availability are: i) accessibility by the end user through both a human-machine interface (HMI) and a machine-to-machine interface (MMI)~\cite{Depuru2011}, and ii) a near real-time sampling frequency~\cite{EuropeanCommission2012}.

\subsubsection*{\ac{sm} penetration}

Having a \ac{sm} is the necessary condition to access the post-metering services' market, which, in turn, relies on the presence of a sufficient number of enabled customers to be able to flourish. This virtuous circle is triggered when the penetration of \acp{sm} allows for the creation of a critical mass of potential users.

\subsubsection*{\ac{sm} standardization}
In some countries, the high volumes of customers to be equipped with a \ac{sm} and the presence of numerous \acp{dso} may make massive rollouts a lengthy and extensive assignment. It is necessary that the proposed solution(s) be standard, certified and tested, and stable in time, in order to foster the interoperability of post-metering technologies. \acp{dso} and \acp{nra} constantly work to this end, testing different solutions, and promoting trials and experimental validations~\cite{VandeKaa2019,Erlinghagen2015}.

\section{International Framework and Literature Review}

With the three pillars defined above as an instrumental guideline, we can further the analysis by presenting a state of the art of post-metering industrial solutions on a global level.

A relatively small number of countries have reached an acceptable penetration rate of \acp{sm} so far. Africa, South America and Central America are still struggling to adopt such technologies and only Brazil is starting to move in this direction thanks to the initiative of major electrical utilities~\cite{WoodMackenziePower2020}. In North America, Canada leads the transition to \ac{ami} (\ac{sm} penetration $>$ 82\%~\cite{Wadhera2019}), while the USA is expected to achieve a penetration of around 80\% by 2025~\cite{WoodMackenziePower2020}. To the best of the authors' knowledge, at present post-metering services are only foreseen as an opportunity for the future, rather than a consolidated reality in both countries. As for the other continents, the Asia Pacific area is very active in the smart metering market. In Oceania, in particular, New Zealand reached 85\% penetration in 2020~\cite{BERG:2020} and Australia will have 7.4 million \acp{sm} by 2025 (penetration rate will be approximately 50\%)~\cite{WoodMackenziePower2020}. In these cases too, there is no clear reference to value-added services or provision of real-time data to users~\cite{ElectricityAuthority-TeManaHiko}. However, East Asia is the leader of the continent: China has completed a first-generation rollout in 2019, South Korea and Japan will complete their rollout in 2022 and 2024 respectively and are currently nearing a 90\% penetration. India is the emerging country and will have 80 million \acp{sm} installed by 2025 (more than 80\% penetration)~\cite{WoodMackenziePower2020,BERG:2020}. In Asian countries, however, second-generation meters are yet to come; post-metering services are not widely disseminated and standardized, to the best of authors' knowledge.

Europe is undoubtedly a pioneer in terms of \ac{sm} technologies, but different member states of the EU have reached different levels of maturity of the technology, both in terms of functionalities supported by \acp{sm} and in terms of penetration. In particular, the two characteristics of data availability mentioned in Section~\ref{sec:def} have been foreseen to be implemented in future installations by several \acp{nra}, but are available on only a limited number of solutions already installed on field. The following states have reached a \ac{sm} penetration of more than 90\% by the end of 2018, according to the European Commission~\cite{Tractebel2019,Tractebel2019a}: Estonia, Finland, Italy, Malta, Spain and Sweden. France will reach this goal by the end of 2021~\cite{CGE:2020}. In Estonia (98.9\% penetration), data are made available to customers through the website of the \ac{dso} with an hourly resolution, but no post-metering services are in place. Finnish second-generation \acp{sm} will offer customers the possibility of accessing meters' data locally in near real-time, but the deployment began only in 2020. Malta has a very high penetration of \acp{sm} based on \ac{plc} (97.3\%) and is installing the second generation of meters based on the Italian technology. Customers will have access to the meters' readings via a web interface. Spain has been deploying two types of \ac{plc}-based \acp{sm} with two different communication protocols, reaching a 99.1\% penetration, but lacks clear standards for the opening of \acp{sm}' data to final customers and third parties. While several solutions have been hypothesized (in-home displays, smart energy boxes, \ac{dso} websites), the granularity of data (1 hour) is still unsatisfactory with respect to the parameter of data availability defined above, and the technology has not been consolidated yet. Sweden has already completed a first-generation \ac{sm} rollout, initiated in 2007, and has planned to install the second generation of \acp{sm} by 2025. The provision of real-time and historical consumption data is defined as a characteristic of the new generation of meters, however there is not a universal standard. In France, the \textit{Linky} \ac{sm} is currently under deployment by Enedis, the French \ac{dso}, and is among the most insightful examples of post-metering activities at present. Linky is able to provide near real-time data to third parties thanks to a device called ERL (``émetteur radio Linky'', French for ``Linky radio transmitter'') that can be installed on the \ac{sm} and uses a two-way radio protocol to communicate (KNX-RF Multi Fast profile, and ZigBee)~\cite{Oudji:2015}. A range of post-metering value-added services can be enabled and several initiatives are ongoing to explore their market potential~\cite{CGE:2020}.

Scientific literature has studied the potential of \acp{han}---which permit communication between the household's internet gateway, home appliances and energy management systems---within a smart grid context. In~\cite{Huq:2010}, authors presented a comparison of different \ac{han} technologies and highlighted the most important factors when it comes to the selection of the technology (such as device ownership, market diversity, interoperability, cost and performance, etc.). In~\cite{Pipattanasomporn:2012}, a \ac{hems} was studied as an effective way to provide demand response services with the use of several appliances. More recently,~\cite{Bazydlo:2018} explored the possibility of reducing peak loads by means of a \ac{han} controller based on a Unified Modeling Language algorithm, and demonstrated its utility and cost efficiency. In~\cite{Yildiz:2017}, authors presented an extensive review of the use of \ac{sm} data, underlying the growing diffusion of \acp{hems} and the difficulty in comparing different experiences due to the diverse characteristics of data.

In Italy, the solution adopted by the \ac{nra} and the \acp{dso} is a notable example thanks to its territorial extent, and both technological maturity and characteristics, such as interoperability standards. Previously published articles analyzed the deployment of first-generation \acp{sm} in Italy within the ``Telegestore'' project~\cite{Botte:2005}, as well as first-generation post-metering technology and its applications in funded projects~\cite{Lombardi2014}. Second-generation Italian \acp{sm} and post-metering services enabled by \chaintwo\@ channel were studied in~\cite{Piti2017} and~\cite{Piti2019}. This paper builds upon such work and presents novel results from \chaintwo\@ experimental campaign, proposes a new classification model for use cases, and investigates the policy and market implications of customer-oriented services in liberalized markets by capitalizing the lessons learned through \chaintwo. The regulatory methodology and the resulting choice of the communication channel are presented in Section IV.1; the technological methodology for analyzing the robustness of \chaintwo\@ and the field results are exposed in Section IV.2; the functional methodology for the study of use cases and its application to the Italian case are illustrated in Section IV.3.

\section{Post-Metering Services in Italy}\label{sec:process}

Italy was the first country in Europe---and one of the first countries worldwide---to initiate a massive installation of first-generation \acp{sm} in the early 2000s as an initiative by E-Distribuzione (the largest Italian \ac{dso})~\cite{Botte:2005}. In 2006, E-Distribuzione's rollout was completed and ARERA (Autorità di Regolazione per Energia Reti e Ambiente, the Italian \ac{nra}), acknowledging the advantages entailed by such an approach, set a target of 95\% \ac{sm} penetration for \ac{lv} users by 2011~\cite{ARERA2006}. In 2016, ARERA set the requirements for the second generation of \acp{sm}~\cite{ARERA2016}, following a public consultation process with all relevant stakeholders.

\subsection{Choice of the Communication Channel}\label{sec:comm}
In defining such requirements, the \ac{nra} decided to dedicate specific efforts to post-metering services, with a focus on the following characteristics: i) post-metering services must be developed by suppliers or third-party service providers designated by customers; ii) the protocol for post-metering services must be standardized and universal in order to foster interoperability; iii) the standards for post-metering communication must follow the needs expressed by market operators. To these ends, ARERA has entrusted the Italian Electrotechnical Committee (Comitato Elettrotecnico Italiano, CEI) with the task of managing a working group for the redaction and periodic revision of technical norms concerning post-metering services. The two versions already issued in 2017 and 2020~\cite{CEI2020,CEI2020a,CEI2020b} have followed public consultations with \acp{dso}, manufacturers, service providers, and other stakeholders.

The standardization of \chaintwo, a communication channel through C-band \ac{plc} between electricity meters and user devices is a clear sign of how a good regulatory approach could bring about significant advantages for all the stakeholders involved. The early collaboration among all the players allowed overcoming some of the barriers that would have been in place if the whole development had been left to the initiative of individual retailers or asset providers. Once the deployment of the second generation of \acp{sm} will be complete, all end users will be able to access the same data set and services by means of a user device, regardless of their \ac{dso}, retailer or location, thus realizing the principle of equality of treatment. \acp{dso} are asked to guarantee that their \acp{sm}' \chaintwo\@ interface be compliant with the CEI regulations. In particular, CEI sets the standards for the data set an end user will receive: in this way, the activation of \chaintwo\@ channel is simplified, because it is not the \ac{dso}'s duty to negotiate the terms of access with service providers and customers. At the same time, device manufacturers and service providers are able to use the same equipment for all the users regardless of the \ac{dso}: this prevents market fragmentation and favors economies of scale.

\begin{figure*}[t]
	\centering
    \subfloat[]{\includegraphics[width=\columnwidth]{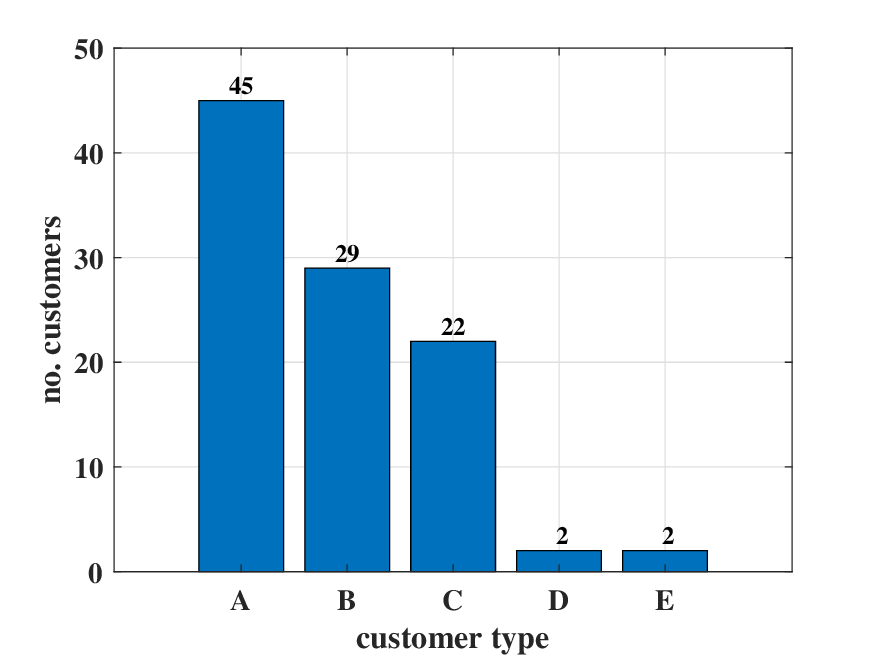}}
    \label{fig:monit_1}
    \hfil
    \subfloat[]{\includegraphics[width=\columnwidth]{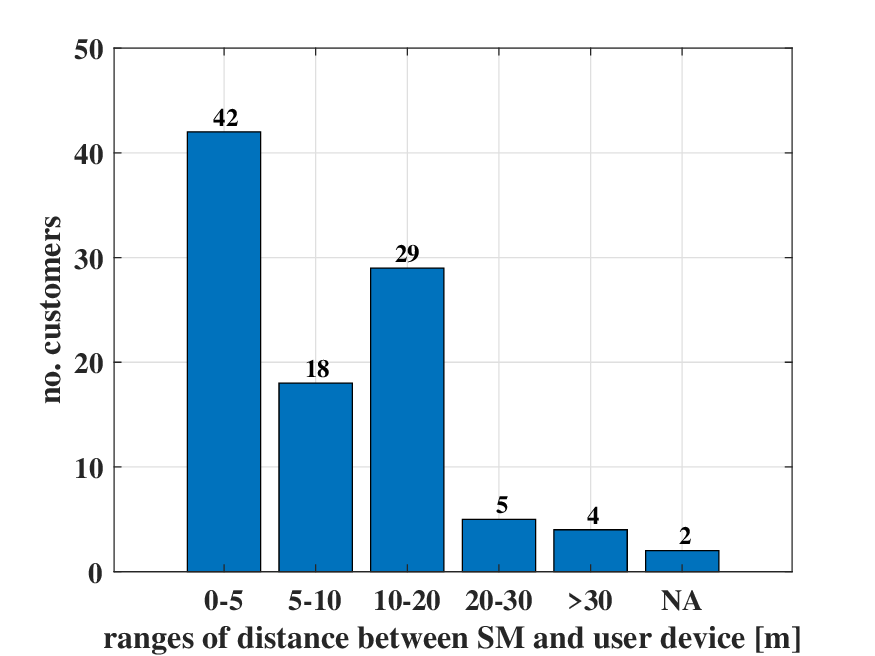}}
    \label{fig:monit_2}
    \caption{Characteristics of the end users in \chaintwo\@ experimental campaign: (a) distribution of the customers according to the type of connection:  A: building; B: independent house; C: apartment block; D: large single-phase \ac{lv} customer; and E: not classified; and (b) distribution of the customers with respect to the distance between the \ac{sm} and the user device.}\label{fig:monit}
\end{figure*}

In this context, the choice of the communication channel for \chaintwo\@ was critical. Among the standards provided by CEI---\ac{plc} on CENELEC C-band (\SIrange{125}{140}{\kilo\hertz}), \ac{rf} @\SI{169}{\mega\hertz} and narrow-band (NB) IoT---the \ac{plc} one presented significant advantages from the point of view of costs and communication efficiency~\cite{Piti2017}. The main three reasons, described in the following, are: i) it is present in all the buildings; ii) it does not require an intermediary or an extra access fee; and iii) it is intrinsically secure. Thanks to the \ac{plc} channel, all households and small businesses' premises are equipped with a local bus enabling an IoT architecture which features the \ac{sm} as an advanced sensor and the user device as an IoT gateway, door to the cloud. Looking specifically at \chaintwo, the SMITP protocol B-PSK technology~\cite{CENELEC2015} enables a channel coded @\SI[per-mode = symbol]{4800}{\bit\per\second}, suitable for low-rate, high-efficiency and low-latency applications. The use of CENELEC C-band guarantees full reliability (see also Section~\ref{sec:campaign}), as this channel's noise level is even lower than CENELEC A-band's, which is currently used by the \acp{dso} for the remote management of \acp{sm} and data collection purposes.

Despite the wide diffusion of radio mobile technology, there are peculiar installation conditions (e.g., under the pavement, in a metal box, in a concrete base, etc.) where the signal of an NB-IoT-based service could be attenuated to inadequate levels. Transferring the IoT connectivity to the user device---and therefore to the end user's apartment or premises---allows providers to have more control on their devices and a better connection in terms of security and continuity of service. Furthermore, the \ac{plc} channel, after the initial capital costs for the development of the solution, is basically free of cost (no access fee is due to any intermediary, as the \ac{dso} itself is the manager of the infrastructure). On the contrary, if a SIM for the management of an NB-IoT-based \chaintwo\@ channel were deemed to be necessary, it would be challenging to determine the cost allocation. Post-metering service providers could indeed manage the installation of a SIM, but, in this case, they would need to operate on \acp{sm}, which are owned by \acp{dso}. On the other hand, if SIMs are provided and installed by the \ac{dso}, the choice about the network operator could cause a ``vendor lock-in effect'': the \ac{dso} would depend on a specific provider, and be unable to switch without substantial cost. E-SIMs installed directly on \acp{sm} could help to avoid this effect. However, the cost for their management---such as the fee for a remote-provisioning platform---would be difficult to allocate or could cause an excessive burden on final users. Again, keeping the user device as an IoT connectivity center permits solving all the above-mentioned problems. Moreover, having defined similar protection mechanisms at an application level, the \ac{plc} communication technology ensures higher security standards than \ac{rf} with respect to man-in-the-middle (MITM) attacks. This technology is particularly suitable for short-range communication, such as that required in \chaintwo\@ applications. An \ac{rf}-based communication channel, given its broadcast nature and long-range reach, could be prone to attacks carried out remotely and from higher distances (more than \SI{100}{\meter}). On the contrary, with \ac{plc}, a malicious attacker would be forced to physically connect to the phase conductor either between the \ac{sm} and the user device or nearby (in a range of up to \SI{100}{\meter} approximately).

The chosen \ac{plc}-based \chaintwo\@ channel finds a suitable framing within the functional reference architecture for communications in smart metering systems as defined by CENELEC in~\cite{CENCENELECETSI:2012}, and represents the H2 interface towards home automation functions.

\subsection{Experimental Campaign and Validation}\label{sec:campaign}
With the objective of monitoring the performances of \chaintwo\@ communication channel, ARERA launched an experiment in April 2017~\cite{ARERA2017}. The experiment was carried out by E-Distribuzione in 2018~\cite{EDistribuzione,Piti2019}. In this phase, it was chosen to focus directly on final customers, in order to take into account the real conditions of installation of user devices in terms of typology of buildings, electrical connections and possible presence of powerline noise on the electric grid. The experiment focused on measuring the actual end-to-end performances of \chaintwo\@ communication channel by comparing the number of messages sent by the \acp{sm} and the number of messages received by the user devices. In compliance with the \ac{nra}'s requirements, the experiment was devised in such a way that could guarantee end users' active participation. Several entities, such as service providers, energy traders and \acp{esco}---generally referred to as asset providers---were allowed to take part on a voluntary basis, provided that they were able to find users and provide them with a user device. A laboratory conformity assessment phase of user devices was considered the enabling condition for the participation. The \acp{sm} used for the experiment were the \ac{lv}, single-phase, second-generation meters produced by E-Distribuzione, aka \textsc{open} Meter. Both passive and active users were involved: the latter were equipped with a \ac{sm} for the measurement of exchanges with the grid (M1) and one for the self-production (M2). Three different message types (aka compact frames) were selected as the scope for the experiment, with the aim of testing messages with different characteristics: i) load curve, namely the active energy absorbed in the last 15 minutes, which is a message sent with a constant frequency of 4 times an hour (T1); ii) instantaneous power, namely the power absorbed upon crossing a band defined as 10\% of contractual power ($P_{\rm n}$), which is a message sent with a high and variable frequency depending on user's consumption profile (i.e., a message is sent whenever the power absorbed crosses one of the thresholds set at $kP_{\rm n}/10$, with $k=1,\dotsc,10$) (T2); iii) exceeding of contractual power or of a predefined consumption threshold (in \si{\kWh}), which is a message sent with a low and variable frequency (T3). Due to the limited number of message typologies examined, it was possible to test only few use cases, mainly belonging to the energy awareness and alarm categories (e.g., A.2 and A.3, as defined in Appendix A). The experiment entailed the activation of \chaintwo\@ channel for 100 final users spread on the Italian territory, different types of buildings and electrical connections (see Fig.~\ref{fig:monit}(a)), in order to guarantee the representativeness of the sample. Seven asset providers were involved with five models of user devices. The distance between the meter and the user device varied significantly (see Fig.~\ref{fig:monit}(b)). RSE, a publicly owned Italian research institute, was appointed as the coordinator and published the results of the experiment~\cite{Tornelli2019}. These results were considered to be satisfactory by all the stakeholders involved: E-Distribuzione, the \ac{nra}, asset providers, and research bodies. Daily rates of success were close to 100\% for all types of messages. In Table~\ref{tab:tab_1}, next page, the results of the experimental campaign are reported.

\begin{table*}[t]
	\centering
    \caption{Summary of \chaintwo\@ experimental campaign results. Three types of messages were analyzed: T1, T2, and T3. The success rate was defined as the ratio between the number of messages received by the user device vs. the number of messages sent by the meter. Both periodic (T1) and event-based messages (T2 and T3) showed excellent performances.}\label{tab:tab_1}
    \includegraphics[width=.9\textwidth]{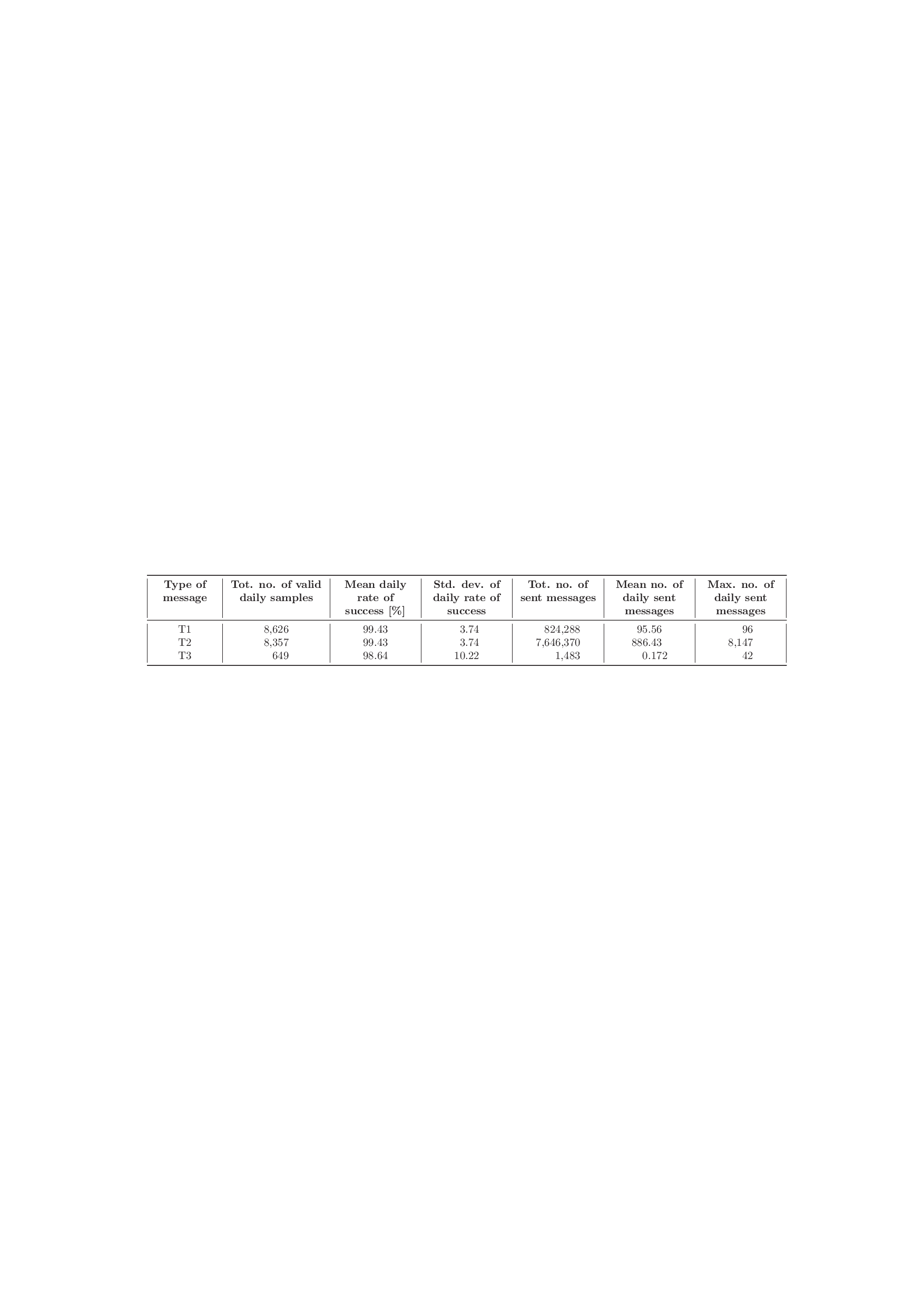}
\end{table*}

As a consequence of the experiment, the \ac{nra} validated \chaintwo\@ communication channel and did not require the development of a back-up channel~\cite{ARERA2020,ARERA2020a}. One of the main takeaways of the experimental campaign was that a reliable technology is a necessary but not sufficient condition to guarantee an effective scale-up of post-metering solutions. In particular, three aspects were found to be crucial in this regard: i) communication among the stakeholders involved should be enhanced and streamlined for quick assistance at any stage of the process; ii) a sound architecture would be required to activate the service quickly upon a customer's request; iii) the whole process should be fully automated in order to ensure operability for large numbers of customers. As a consequence, E-Distribuzione elaborated a complex digital system entirely dedicated to value-added services, a work that culminated in the launch of a \chaintwo\@ web portal, in March 2019. Through this channel, asset providers are allowed to verify in real time whether a \ac{pod} is eligible to communicate via \chaintwo, i.e., whether a \ac{pod} is supplied by E-Distribuzione and is equipped with an \textsc{open} Meter. Subsequently, service providers are entitled to create a \ac{pod}-user device pairing to initiate the communication: such operation is remotely managed and performed within few hours. The presence of a web interface was key to guaranteeing an effective scale-up of the solution and prompt intervention in case of faults.

Despite the work done by ARERA and E-Distribuzione, the main weakness of \chaintwo\@ is currently the relatively small number of active users (in the order of magnitude of hundreds). The reasons why market operators have not reacted in a prompt way are mainly two: i) technical standards have already been amended once, which implied that all the stakeholders had to modify the hardware and software developed for the experimental campaign in order to make them compliant with the new norms; and ii) given the highly innovative character of the solution, greater effort would have been required from all parties, since the beginning of the project, to communicate the potential impacts \chaintwo\@ could have on retailers, service providers and, ultimately, end users. Furthermore, the outbreak of COVID-19 might have had significant repercussions on the dissemination of \chaintwo.

E-Distribuzione has set the pace for the implementation of this technology and initiated the massive rollout of second-generation \acp{sm} in 2017. Following this trace, other \acp{dso} have already decided to adopt the \textsc{open} Meter designed by E-Distribuzione: \chaintwo\@ is currently potentially available to more than 20 million customers (and counting), and is planned to reach all Italian \ac{lv} users by the end of 2026.

\subsection{Use Cases for Post-Metering Services}\label{sec:usecases}
By opening a new communication channel towards the customer, a new paradigm is introduced in the final user's universe. Value-added services become available to everyone independently of the specific will and interest of the \ac{dso} that is meant to enable them. The range of customers interested by these services is quite diverse: small businesses---especially supermarkets, bank branches, post offices, insurance agencies, industrial offices, restaurants---as well as households; consumers as well as producers and prosumers; individual users as well as energy communities. For all these categories, it is crucial to be able to make use of cost-effective, universal and scalable solutions: customers must be in the condition of pulling the market and setting the standards for the services they need.

\begin{table*}[t]
	\centering
    \caption{Identification of potential use cases based on the $\rm I^2MA$ model. Each row represents one of the four levels of service, while the columns identify the benefits. In each cell, one or more use cases are reported.}\label{tab:usecases}
    \includegraphics[width=.9\textwidth]{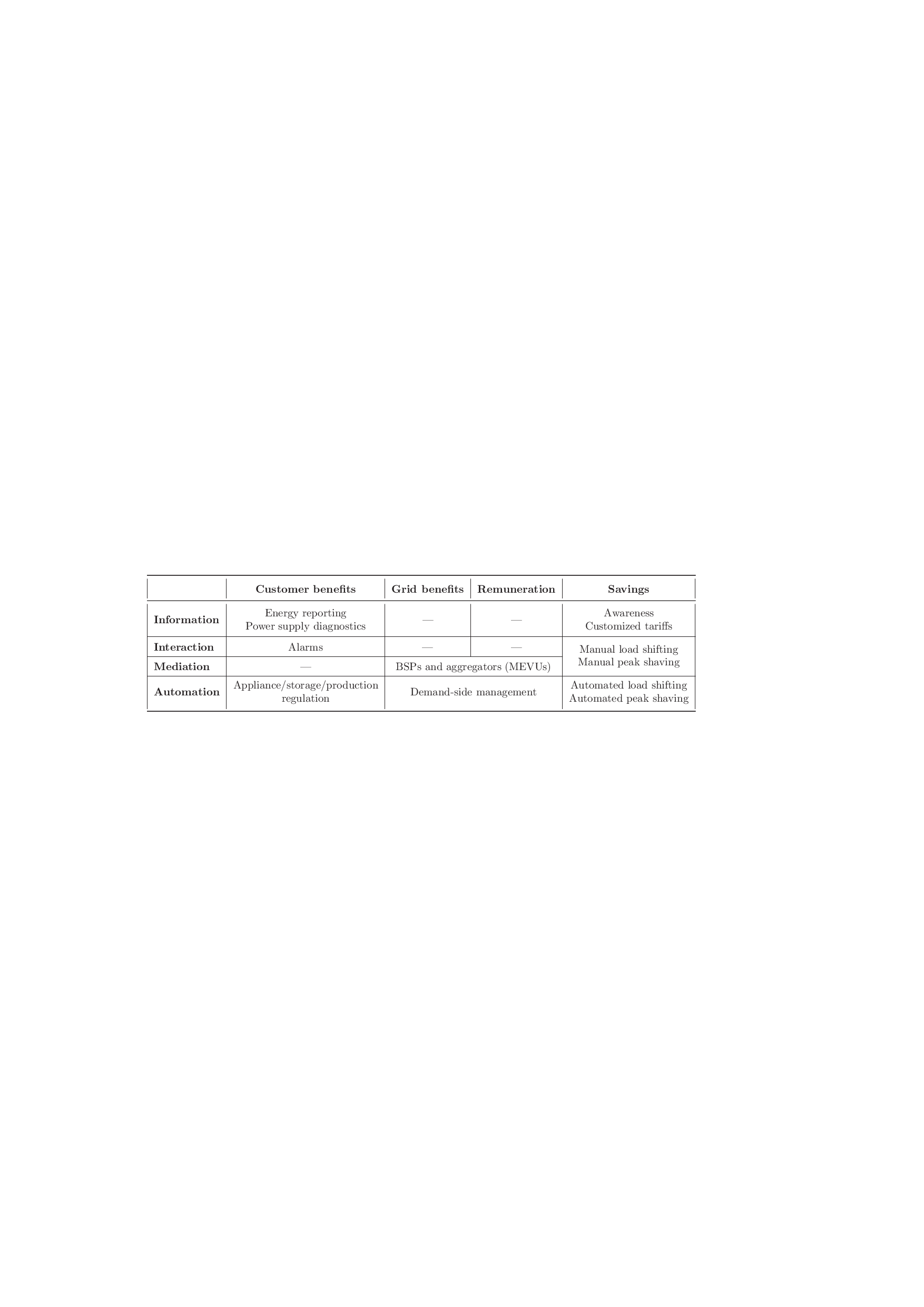}
\end{table*}

Thanks to an early \ac{sm} rollout, Italy has gained valuable expertise in providing end customers with near real-time metering data. The first prototype of post-metering services, Enel Info+, was launched as an experiment by E-Distribuzione in the city of Isernia. The communication channel chosen was based on PLC A-band (aka \chainone) and the solution featured an in-home device provided by E-Distribuzione, which came together with a small display. The experiment led to the creation of a first generation of in-home devices, the so-called Smart Info, that can be considered an experimental DSO-produced precursor of a \chaintwo\@ user device. This experience paved the way for similar projects in several Italian regions. As stakeholders' expertise grew, it became more and more evident that limiting post-metering services to a small number of use cases, mainly for domestic customers and for the purposes of energy awareness, would be a short-sighted choice. In the working groups that led to the adoption of \chaintwo, a variety of opportunities have been explored and several use cases have been devised. The remaining part of this paragraph aims at presenting an overview of value-added services enabled by \chaintwo. In order to do so, we propose a classification model called $\rm I^2MA$ based on four levels of service characterized by a growing technological and market maturity: \emph{Information}, \emph{Interaction}, \emph{Mediation}, and \emph{Automation}. Each level is matched with a set of benefits for the system and/or the end user (as briefly reported in Table~\ref{tab:usecases}, next page) with the aim of obtaining the use cases detailed in Appendix A and outlined in the following.

The first level of service is the displaying of information, which allows users to access consumption and production data, when available. This is the basic use case of energy awareness, that creates the possibility for the customer to continuously monitor their behavior and expenses~\cite{Carroll2013}. The previously mentioned Enel Info+ experiment helped confirm the usefulness of energy awareness tools in order to obtain savings: more than 60\% of customers observed bill reductions after the installation of a Smart Info~\cite{Lombardi2014}.  Displaying data also enables the use case of energy reporting, which is particularly advantageous in the cases of energy intensive businesses. Italian legislation, for instance, has introduced the concept of compulsory energy reporting for specific categories of companies, such as big companies or energy-intensive businesses since 2016~\cite{Legge102}. On an annual basis, companies must report all savings stemming from improvements in the productive cycle, but also from behavioral changes. The use of near real-time electricity consumption data can be of great importance for ensuring reliability and accountability of such claims. This approach also extends the possibility of energy saving interventions to small businesses and the tertiary sector. Economic savings can also be obtained thanks to the new tariff structures enabled by second-generation \acp{sm}, that allow retailers to provide offers based on customers' consumption behavior. Users can also avoid extra costs due to the oversizing of the supply, by monitoring their energy consumption through \chaintwo\@ channel. The possibility of running diagnostics on power supply quality (e.g., interruptions, curtailments, voltage and frequency variations, etc.) represents an additional benefit for end users.

\begin{figure*}[t]
\centering
\includegraphics[width=.9\textwidth]{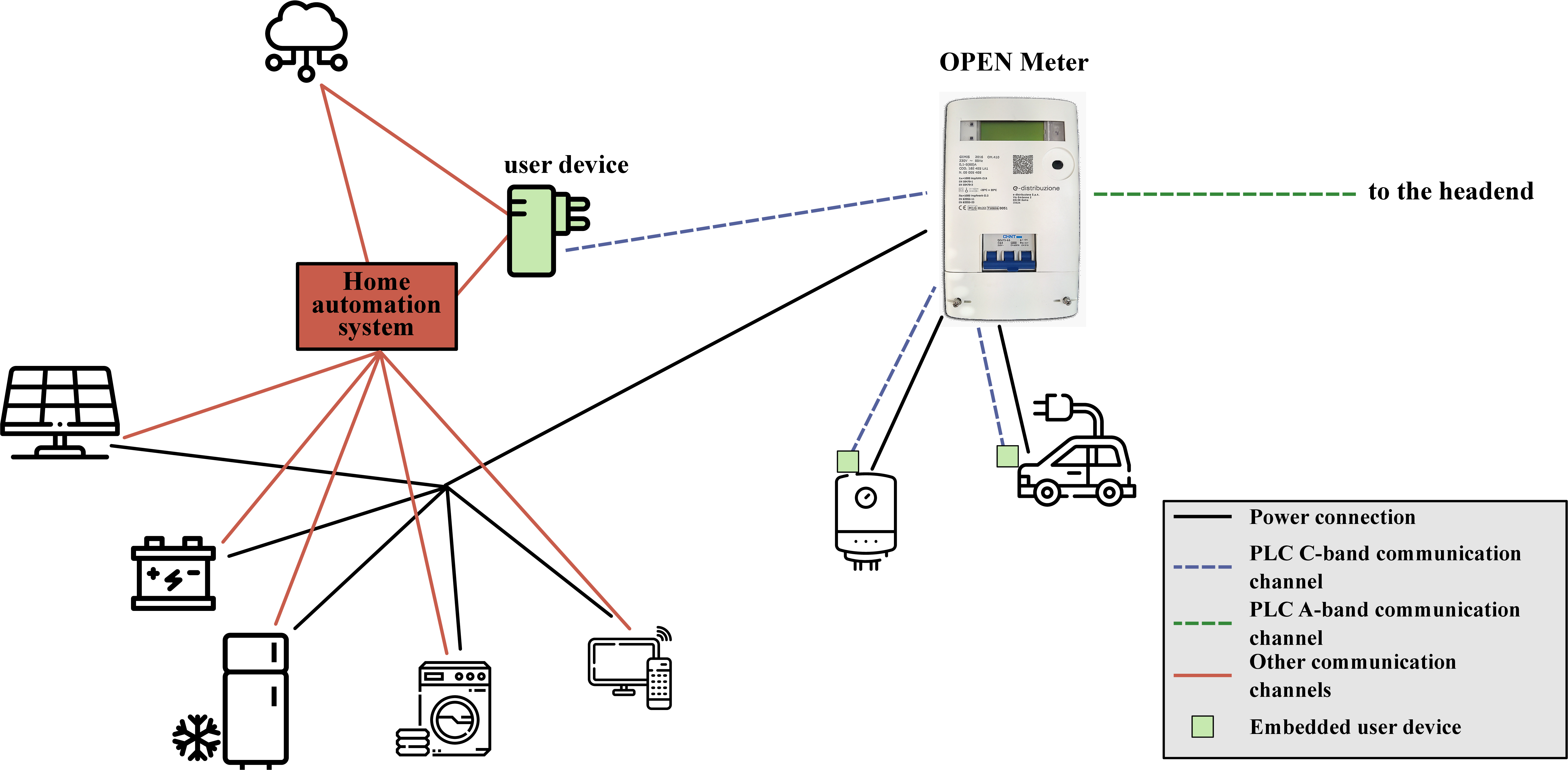}
\caption{Schematic diagram of a home automation system enabled by \chaintwo. Physical and communication layers are represented, and the user device acts as an IoT connectivity center.}
\label{fig:it_case}
\end{figure*}

The second level of service is interaction. Coming to know their energy consumption, customers are able to roughly understand which are the most energy-intensive appliances, spot failures or losses and do a basic manual load shifting of the most power-consuming equipment to periods of the day when electricity costs less, according to the time-of-use tariff structure of the contract. Furthermore, it is possible to set alarms in order to notify when the user exceeds a pre-determined threshold of energy consumption, as well as avoid overloads and consequent power cutoffs. This use case is particularly interesting in the Italian context, where a switch-off algorithm is in place: feeding is cut within a time that is inversely proportional to the excess power absorbed. A prototype of such functionalities was already present in the Enel Info+ experimental project and was above all appreciated by customers above all, however, near real-time data are necessary to make this use case fully accessible and convenient.

The third level of service is the opening of data to third-party mediators. In this context, the April 2020 revision of CEI technical norms defined the role of \chaintwo\@ with respect to balancing service providers (BSPs) and aggregators, which can enable the participation of final users to the dispatching services' market. \chaintwo\@ user devices are able to guarantee that the data be frequent enough to comply with the requirements of the \ac{tso}, which organizes the market. Users can be gathered in \acp{mevu}\footnote{The interested reader may look at the original documentation, in Italian, where \acp{mevu} are called ``unità virtuali abilitate miste (UVAM).''}, clusters that include consumption, production and storage systems, and are allowed to provide tertiary reserve, congestion resolution and balancing services~\cite{Marchisio2020}. Users are paid not only for the energy they produce---or do not consume---when required to do so, but also for the power they make available at all times (i.e., the modulation of power they can guarantee at any given moment, when the \ac{tso} requires it). It is a win-win result, because both the grid and the user benefit from it. This use case can be activated manually, if human intervention is required for switching on/off those loads that contribute to provide such services to the grid, or automatically.

Automation represents the fourth level of service, opening up several use cases and encompassing most of the features of the previous levels.. Automatic load shifting permits using the most energy intensive appliances in the moments of the day when the price of electricity is the lowest. In this case, the user device becomes a key element for the management of smart appliances and the rational exploitation of energy storage systems. At the same time, smart appliances can be set so as not to cause power cutoffs and can be modulated or dimmed for the consumer's benefit. An example of this scenario is the case of an \ac{ev} charging point connected to the domestic network. Two different needs coexist: charging the \ac{ev} battery and using other energy-intensive household appliances (e.g., oven, washing machine, dishwasher, etc.), some of which have nonlinear load profiles. Similar conditions occur when electric boilers or HVAC systems are used in domestic environments or small businesses. The transmission of near real-time data to a user device embedded in the appliance allows creating algorithms for their smart utilization and goes a step further the basic scenario of machine-user interaction facilitated by specific alarms. The potential of this use case can be enhanced through the integration of non-intrusive load monitoring (NILM) algorithms with the purpose of disaggregating load curves, managing appliances, and optimizing loads and storage systems with machine learning methodologies. \chaintwo, by enabling post-metering services, can ultimately become the backbone of a smart home (an example of which is represented in Fig.~\ref{fig:it_case}) and be integrated into an \ac{han}.

Residential demand-side management is expected to soar in this decade and account for 40\% of the increased potential of demand-side management by 2030~\cite[p.~248]{IEAWEO:2020}: in this context, the \ac{mevu} project, integrated with \chaintwo\@ can be a valuable piece of the puzzle.

The experience of Enel Info+ and similar projects (which involved around \num{50000} users in total) can be legitimately considered as a minimum viable product (MVP) for the development of the value-added services enabled by \chaintwo. The interest that these initiatives have sparked, despite their numerous limitations and shortcomings, is a proof of the receptivity of the market to post-metering services. The growing involvement of citizens in shaping the future of electrical networks in Italy is witnessed by the recent enactment of regulations for local energy communities, which led to the creation of 15 pilot projects with the involvement of academic and industrial entities~\cite{RSE}. \chaintwo\@ has proven to be a reliable communication channel and has the potential to provide accurate near real-time measurements to incentivize, track and optimize renewable energy production. Whether \chaintwo\@ will be able to take a quantum leap will depend not only on final customers, but also on the ability of the most visible players to give it full visibility and communicate its potential.

\section{Lessons Learned and General Recommendations}
\begin{figure*}[t]
	\centering
    \includegraphics[width=.9\textwidth]{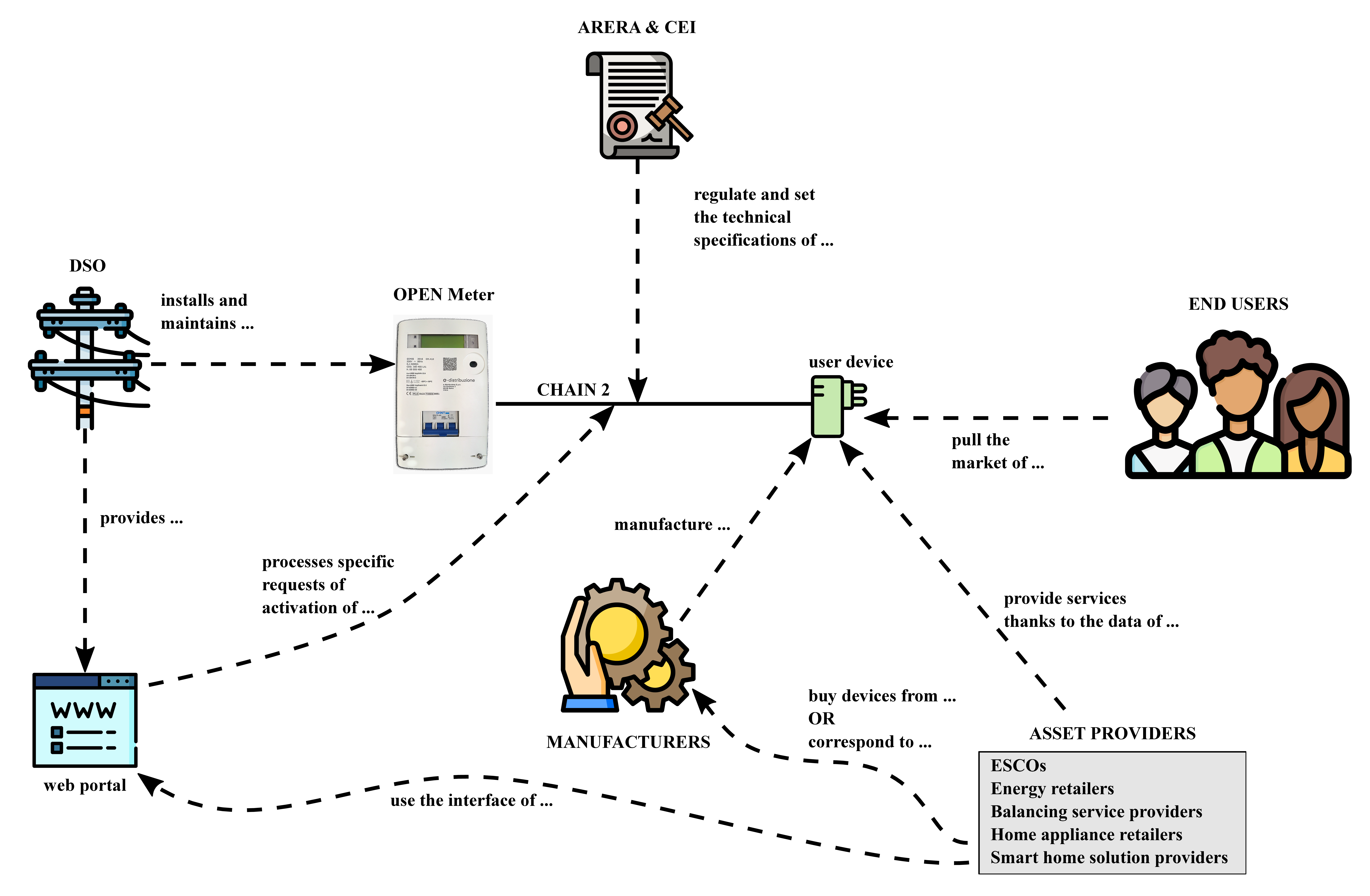}
    \caption{Market structure of \chaintwo. The focal point is represented by \chaintwo\@ channel, which connects the \textsc{open} Meter to the user device. Stakeholders are represented around this cluster. Arrows show relationships among stakeholders and/or system components.}\label{fig:marketstructure}
\end{figure*}

Industry experts and researchers may benefit from the experience of \chaintwo, whose market structure is represented in Fig.~\ref{fig:marketstructure}, next page. Some instrumental recommendations can be derived based on the lessons learned by the authors. Such recommendations are not country-specific and can be extended to any liberalized electricity market. It is worth mentioning that the market of post-metering value-added services in France, which has attained a similar maturity level, has led experts to draw similar conclusions~\cite{CGE:2020}.

\begin{enumerate}
  \item The role of \acp{dso} is to provide a stable environment for third parties and support them in the creation of competitive business models in the market of post-metering services. In other words, \acp{dso} are entrusted with the task of providing an enabling technology which gives end users the ability to pull the market. Therefore, post-metering value-added services are transparent to \acp{dso} and emerge from market competition and interaction between customers and service providers.
  \item In exchange, \acp{dso} must take the responsibility of certifying measurements, without which all the possible post-metering services are likely to be confined to the user's premises and not be exploited by higher-level stakeholders (such as aggregators, \acp{esco}, \acp{tso}, etc.). Furthermore, standardization plays a key role in avoiding these services from becoming a prerogative of a niche of tech-savvy consumers, thus crippling the potential growth, and the consequent beneficial impacts, of this market.
  \item While post-metering value-added services should be user-driven, technological innovation should rather be supported by regulation. In this context, \acp{nra}, by acting as need collectors, allow all the stakeholders to be involved in the definition of the features of such technological framework. In line with this objective, \acp{nra} are called to create working groups that gather representatives from companies (belonging to all the categories mentioned in Section~\ref{sec:campaign} and \ref{sec:usecases}), consumers, \acp{dso}, etc. Periodically, the group should report on national and international trends on post-metering services and highlight new needs and opportunities.
  \item High-level players with visibility and credibility must advertise this type of value-added services by acting as a sounding board rather than merely setting up the enabling context. While the market needs to develop to be self-standing, raising awareness about the existence of a window of opportunity---for example by organizing public events for energy players and the general public---might spur entrepreneurship and industrial initiative. At the same time, citizens must become aware of their individual and collective responsibility in being part of this technological change, and understand its impact in the context of the energy transition.
\end{enumerate}

\begin{table*}[b]
	\centering
    \caption{Application of the $\rm I^2MA$ model to the Italian use cases. Each row represents one of the four levels of service, while the columns identify the benefits. The model allows classifying and properly framing the use cases defined in~\cite{CEI2020}.}\label{tab:usecases_ita}
    \includegraphics[width=.9\textwidth]{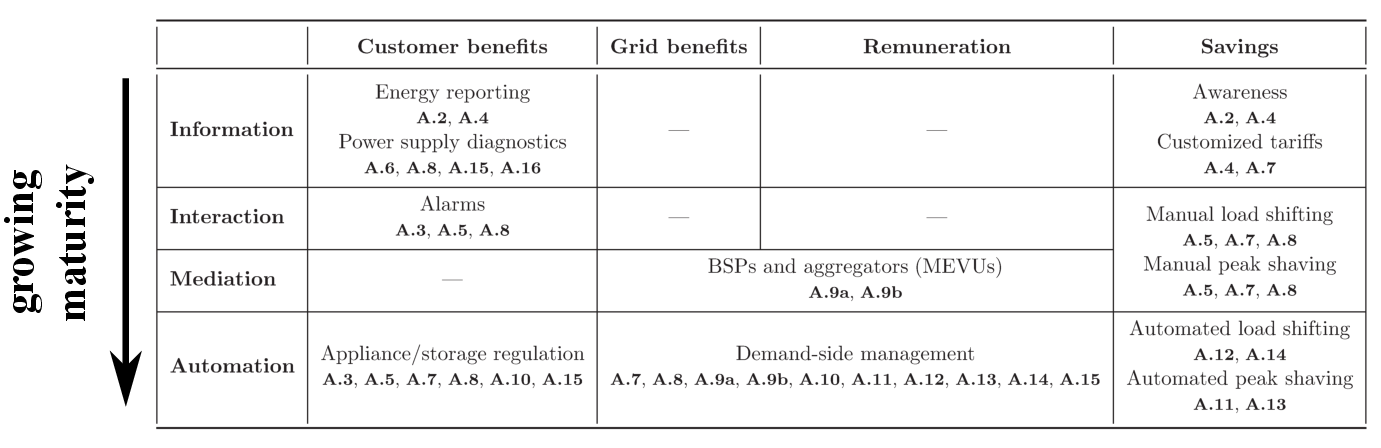}
\end{table*}

\section{Conclusion}
This paper described the framework of post-metering value-added services for \ac{lv} customers, highlighting the key role played by the \ac{ami} and the \ac{sm}. The Italian case was presented as a notable example of a successful process of selection of a suitable technology, its implementation and the use cases enabled. In particular, in the early 2000s E-Distribuzione voluntarily deployed the first generation of \acp{sm} and was soon followed by other Italian \acp{dso}. After the capitalization of almost 15 years of field operation and thanks to the instrumental cooperation between public and private sectors, the Italian \ac{nra} called for the definition of a new generation of \acp{sm}. During this process, the need for the standardization of a communication channel towards end users emerged. The choice of involving all the different stakeholders was key to having a broad view on the issue. \chaintwo, the resulting outcome, was a satisfactory synthesis of the needs and requirements of the many players involved. Its performance rates ($>95\%$) were assessed through an experimental campaign and were found to exceed stakeholders' expectations. The analysis of the strong and weak points of this experience allowed authors to define some good practices and recommendations, especially shedding light on the ways to maximize the efficiency of decision processes and clarifying the role of each player involved. A classification model for post-metering use cases (which we called $\rm I^2MA$) was also proposed and applied to the Italian case resulting in 16 use cases, each indicating the stakeholders involved, the maturity level, and a set of possible service providers. Future research will have the opportunity to explore the market potential of each of these services.

Not only does a technology like \chaintwo\@ envisage a smart grid promising to enable advanced use cases, e.g., demand-side management, but it also makes it tangible for customers by realizing the data exchange in an effective and reliable manner. Thus, it becomes the starting point for setting up customer-oriented solutions that are in line with the new role of the \ac{dso} as a market facilitator, enabler of new business models, and promoter of the energy transition. Post-metering technologies ensure the creation of value for customers, electric systems and, ultimately, for the environment as well.

\textit{This research did not receive any specific grant from funding agencies in the public, commercial, or not-for-profit sectors.}

\setcounter{table}{0}
\appendix{Application of the $\rm I^2MA$ model to the use cases defined in~\cite{CEI2020}.}

In 2020, following the request by ARERA, the Italian Electrotechnical Committee issued a technical standard series\footnote{One of the authors of the present paper (i.e., Daniele Mardero) actively contributed to the working group that led to the issuing of this series.} for regulating the communication between \acp{sm} and user devices. Part 1 of this series is dedicated to use cases~\cite{CEI2020}. While the $\rm I^2MA$ model described in Section 4.3 allowed us to identify some high-level use cases (see Table~\ref{tab:usecases}), in this Appendix we are applying the same model to derive the specific operational use cases for the Italian architecture, as reported in Table~\ref{tab:usecases_ita}. Moreover, for each use case we define the maturity level, the potential service providers, and whether or not it is a smart home enabler. Some use cases may have multiple levels of maturity, as they have multiple applications (see Table~\ref{tab:usecases_ita} for further details).

\textbf{A.1a} and \textbf{A.1b}: \textbf{Installation, activation of the service and configuration of the user device}

These use cases encompass the initialization procedure, the activation of \chaintwo\@ and the configuration of the \textsc{open} Meter and the user device. \emph{\textit{Maturity level:}} NA. \emph{\textit{Service providers:}} NA. \emph{\textit{Smart home enabler:}} NA. \textit{Nnote: these use cases have the sole purpose of activating the user device and are not present in Table~\ref{tab:usecases_ita}.}

\textbf{A.2}: \textbf{View of consumption, and energy and power production}

Provide a service that enables the customer to view or manage consumption, production and feed-in of electricity. \emph{\textit{Maturity level:}} low. \emph{\textit{Service providers:}} energy retailers, \acp{esco}, home appliance retailers. \emph{\textit{Smart home enabler:}} no.

\textbf{A.3}: \textbf{Warning for exceeding the available power and possible intervention of the circuit breaker}

Provide a service that warns the customer when the available power limit is exceeded and, if need be, informs them about the remaining time before the intervention of the circuit breaker, in order to be able to counteract manually or automatically. \emph{\textit{Maturity level:}} medium/high. \emph{\textit{Service providers:}} energy retailers, \acp{esco}, home appliance retailers, balancing service providers. \emph{\textit{Smart home enabler:}} potentially.

\textbf{A.4}: \textbf{View of consumption with estimate of cost and, for prosumer customers, view of generated power and estimate of income}

Provide a service that enables the customer to view consumption, production, feed-in of electricity and a view of the instantaneous trend of power flows. It can use the embedded functionalities of smart devices (such as appliances or smart plugs) to provide detailed disaggregated consumption data. It includes the ability to provide an estimate of utility bill costs and anticipated income for the local generation plant. Through a user application, it is also possible to view some of the contractual information, as defined by the retailer (e.g., the customer verifies whether a contract change has already been made). The consumption curve can be analyzed \textit{a posteriori} to warn the customer about deviations that could be indicators of problems, and the service can provide suggestions to maximize the efficiency and the savings. \emph{\textit{Maturity level:}} low. \emph{\textit{Service providers:}} energy retailers, \acp{esco}, home appliance retailers. \emph{\textit{Smart home enabler:}} no.

\textbf{A.5}: \textbf{Warning for exceeding a power limit set by the customer}

The user interface warns the customer that the power limit they had previously set in the user device or in the Cloud, has been exceeded. The goal is to limit the consumption or verify the starting of an appliance, to apply parental controls, etc. The limit is chosen by the customer among predefined values. \emph{\textit{Maturity level:}} medium/high. \emph{\textit{Service providers:}} energy retailers, \acp{esco}, home appliance retailers, balancing service providers. \emph{\textit{Smart home enabler:}} potentially.

\textbf{A.6}: \textbf{Information on power supply interruptions}

Provide a service that informs the customer about the power supply interruption \textit{a posteriori} both for power outages and for planned maintenance interventions. \emph{\textit{Maturity level:}} low. \emph{\textit{Service providers:}} energy retailers, \acp{esco}, home appliance retailers. \emph{\textit{Smart home enabler:}} no.

\textbf{A.7}: \textbf{Innovative contract types enabled by smart device (e.g., dynamic energy pricing, variable contractual power and pre-paid consumption)}

In the hypothesis of regulating the possibility of accessing a customized set of tariffs, the customer is promptly informed and can intervene either manually or through an automatic coordination system in order to better manage their energy consumption. \emph{\textit{Maturity level:}} low/medium/high. \emph{\textit{Service providers:}} energy retailers, balancing service providers. \emph{\textit{Smart home enabler:}} potentially.

\textbf{A.8}: \textbf{Emergency limitation of exchanged active power}

Intervention of the circuit breaker in case of temporary limitation of active power withdrawn from / fed into the grid by the \textsc{open} Meter. Event notification for temporary limitation of available power and, in case of exceeding this limitation, warning about imminent intervention of the circuit breaker, in order to reduce power demand peaks in an emergency condition. \emph{\textit{Maturity level:}} medium/high. \emph{\textit{Service providers:}} energy retailers, \acp{esco}, home appliance retailers, balancing service providers. \emph{\textit{Smart home enabler:}} potentially.

\textbf{A.9a}: \textbf{Market demand response (for power limitation)}

Limitation of active power withdrawn from / fed into the grid, as requested by the system provider (e.g., aggregator) with the aim of limiting power demand peaks. The user interface warns the customer about exceeding the power limit imposed in that moment within the framework of a demand response program. \emph{\textit{Maturity level:}} high. \emph{\textit{Service providers:}} home appliance retailers, balancing service providers, smart home solution providers, \ac{ev} charging station suppliers. \emph{\textit{Smart home enabler:}} yes.

\textbf{A.9b}: \textbf{Participation in the dispatching services' market, also by means of an aggregator}

Provide support to the network by participating in the dispatching services' market, under appropriate remuneration. The service can be supplied directly by the customer identified as dispatching user or through an entity called aggregator, whose task is to coordinate consumption and production of more customers. \emph{\textit{Maturity level:}} high. \emph{\textit{Service providers:}} home appliance retailers, balancing service providers, smart home solution providers, \ac{ev} charging station suppliers. \emph{\textit{Smart home enabler:}} yes.

\textbf{A.10}: \textbf{Scheduled start of a smart appliance}

Scheduled start of a smart appliance based on the cost of electricity, the availability of power generated by the local generation plant, and the difference between the available power and the power withdrawn. \emph{\textit{Maturity level:}} high. \emph{\textit{Service providers:}} home appliance retailers, balancing service providers, smart home solution providers, \ac{ev} charging station suppliers. \textit{\textit{Smart home enabler:}} yes.

\textbf{A.11}: \textbf{Peak shaving with storage}

Use of storage and local generation plant to cut power demand peaks and reduce additional expenditure. \emph{\textit{Maturity level:}} high. \emph{\textit{Service providers:}} home appliance retailers, balancing service providers, smart home solution providers, \ac{ev} charging station suppliers. \emph{\textit{Smart home enabler:}} yes.

\textbf{A.12}: \textbf{Load shifting with storage}

Use of storage and local generation plant to withdraw power from the grid during the time periods when electricity is cheapest. \emph{\textit{Maturity level:}} high. \emph{\textit{Service providers:}} home appliance retailers, balancing service providers, smart home solution providers, \ac{ev} charging station suppliers. \emph{\textit{Smart home enabler:}} yes.

\textbf{A.13}: \textbf{Peak shaving with smart appliances}

Smart appliance management to cut power demand peaks and reduce additional expenditure. \emph{\textit{Maturity level:}} high. \emph{\textit{Service providers:}} home appliance retailers, balancing service providers, smart home solution providers, \ac{ev} charging station suppliers. \emph{\textit{Smart home enabler:}} yes.

\textbf{A.14}: \textbf{Load shifting with smart appliances}

Smart appliance management to withdraw power from the grid during the time periods when electricity is cheapest. \emph{\textit{Maturity level:}} high. \emph{\textit{Service providers:}} home appliance retailers, balancing service providers, smart home solution providers, \ac{ev} charging station suppliers. \emph{\textit{Smart home enabler:}} yes.

\textbf{A.15}: \textbf{Power diagnostics of a smart appliance}

Allow the smart appliance to know the status and the quality of the electricity supply, at the present moment or in the recent past, in order to enable advanced diagnostic functions. \emph{\textit{Maturity level:}} medium. \emph{\textit{Service providers:}} energy retailers, home appliance retailers, smart home solution providers. \emph{\textit{Smart home enabler:}} yes.

\textbf{A.16}: \textbf{Power supply quality check}

Provide the user device with the ability to perform simple power quality checks (e.g., monitoring interruptions, voltage quality, etc.) at the present moment and in the recent past. \emph{\textit{Maturity level:}} low. \emph{\textit{Service providers:}} energy retailers, \acp{esco}, home appliance retailers. \emph{\textit{Smart home enabler:}} no.

%------------------------------------------------------------------------------------
%\IEEEtriggeratref{63}
%\balance

\end{document}